\documentclass[twocolumn,showpacs,aps,floatfix]{revtex4}
\usepackage{graphicx}
\usepackage{amsmath}

\begin{document}
\title{Relativistic configuration-interaction calculations of U I hyperfine constants}
\author{Igor M. Savukov}
\date{July 2020}

\begin{abstract}
Neutral uranium (U I) is a very difficult atom for theoretical calculations due to a large number of valence electrons, six, strong valence-valence and valence-core correlations, high density of states, and relativistic effects. Configuration-interaction many-body perturbation theory (CI-MBPT) can treat efficiently valence-core correlations and relativistic effects, but because the formalism was developed for Dirac-Hartree-Fock (DHF) starting potential that does not contain valence electrons, quite large CI space is needed to compensate for +6 charge of such a potential. Much more efficient  is  relativistic configuration-interaction (RCI) approach which uses relatively accurate starting DHF potential that includes some valence electrons to make the valence electron Hamiltonian diagonally dominated for some states. Here we report calculations of U I hyperfine constants of several low-energy states using the RCI method with the starting potential that includes four f valence electrons. With this starting potential, it is possible to use the single-configuration approximation or small basis sets to obtain quite accurate results for hyperfine structure constants. In fact, by scaling nuclear magnetic moment, the agreement for 5 levels was within 5\%, and a new magnetic moment can be recommended 0.43(2). The method can be further developed to include more extensive data sets to improve accuracy and can be applied to other atoms and for calculations of other properties, for example, relevant to fundamental symmetry tests.         
\end{abstract}


\maketitle
\section{Introduction}
Development of atomic theory for complex atoms, U I in particular, is needed because of  global security \cite{Liu02,Quentmeier01,Lebedev}, atomic energy \cite{Campbell17}, and many basic science applications. For example, spectroscopy of the U atom can be used to detect uranium and characterize its isotopic content for nuclear forensics \cite{Liu02,Quentmeier01,Lebedev} and treaty monitoring. Measurements in radio-active isotopes, especially  minor isotopes, are complicated by  safety restrictions and limited availability of materials, while hyperfine structure knowledge would help to estimate the sensitivity of detection and understand spectroscopic features, especially when high-resolution methods are employed. Also if the possibility of trapping of atoms has to be evaluated, for example for ultra-sensitive detection methods, it is necessary to know the hyperfine structure for multiple levels.   In fundamental physics, atomic structure theory can be applied to calculations of atomic electric dipole moment (EDM) for the interpretation of EDM measurements,  axion-induced effects in atoms \cite{Stadnik}, and alpha-constant variation \cite{Uzan,Flambaumalpha,Savukovalpha}. Uranium isotopes are also of interest for measuring the stellar ages \cite{Cayrel}

However, theory of the uranium atom is quite challenging due to a large number of valence electrons, strong mixing between states, strong valence-core interaction, and relativistic effects. While theoretical energies can be brought into agreement with adjustable parameters, as it is routinely done on the basis of Cowan's code \cite{Cowan,Guyon}, this does not guarantee the correctness of the wavefunctions, which have to be tested.  Transition probabilities can be used to test the wavefunctions, but their experimental values have significant uncertainty \cite{Henrion,Bieniewski,CorlissUI}, which arises from uncertainty of atomic density measurements, local thermodynamic equilibrium (LTE) assumption, or photo-detector calibration.  Landé $g$-factors, which are much more accurately measured, give useful information for testing purity of configurations and determination of L and S of the wavefunction, at least in non-relativistic LS coupling approximation, valuable for the state identification, but $g$-factors do not depend on radial wavefunctions. Hyperfine constants, on the other hand, can be measured with good precision \cite{Hackel, Avril, AvrilTh} and are sensitive to behaviour of radial wavefunctions near the nucleus and relativistic effects \cite{AvrilTh}, so comparison between theory and experiment can provide test of the wavefunctions. Additional motivation for hyperfine constant measurement has been applications in Atomic Vapor Laser Isotope Separation (AVLIS) project \cite{Avlis}.  

In terms of fundamental symmetry applications, such as EDM \cite{SavukovEDM, DzubaEDM, PorsevEDM}, axion-search calculations \cite{Stadnik}, or alpha-variation search \cite{Uzan,Flambaumalpha,Savukovalpha} the result depends strongly on behaviour of wavefunction near the nucleus and/or relativistic effects, which can be tested by comparison of theoretical and experimental  hyperfine constants.  

There are several theoretical approaches that can be used for  calculations in complex atoms, such as U I. The most common are approaches based on Cowan's code. While the orbitals can be obtained quite optimal because the non-relativistic potential includes all valence electrons self-consistently, among multiple relativistic effects only the spin-orbit term is included. Also correlation corrections with the core are difficult to include. With introduction of a large number of scaling parameters, quite accurate energy levels and $g$-factors can be obtained to identify and classify observed levels \cite{Petit}; however, this approach does not guarantee high quality of wavefunctions. To improve the wavefunctions and transition data accuracy,  polarization potentials were introduced to achieve reasonable agreement with experiment: in La II \cite{KulagaEgger} and U II  \cite{Gamrath} for strong lines. In case of La II, it appears that  better accuracy is achieved using configuration-interaction many-body perturbation theory (CI-MBPT) approach with a relatively small number of adjustable parameters to scale the second-order MBPT to improve the accuracy of valence-core interactions \cite{SavukovLaII}. 

However, many potentially valuable methods beyond relativistic Hartree-Fock (Cowan's code) model were not tested in U I.    Among them, relativistic configuration interaction (RCI) method is an accurate all-order approach for treating valence interactions, but with the number of valence electrons increasing, the CI matrix grows exponentially too large  to include highly excited states. Thus the number of basis set radial orbitals is reduced with loss of precision. Furthermore the frozen-core approximation is not sufficient and valence-core correlations have to be treated, as it was evident in case of La II calculations \cite{SavukovLaII}. In case of heavy atoms, these correlations are not small and second-order perturbation theory, used in an amended method, CI-MBPT is insufficient. One solution is to introduce multiple factors to scale the second-order corrections, and it has been shown that agreement with experiment for energies is substantially improved, for example in case of U III and Th I, allowing  matching theoretical and experimental levels \cite{SavukovU+2,SavukovThI}. Another solution is to replace in 
CI-MBPT the second-order MBPT with all-order MBPT corrections, in the so-called  CI-all-order method \cite{SavukovU+2}.

In addition to correlations, relativistic effects are significant too. Relativistic CI-MBPT approach is almost ideal, except that some relativistic corrections, such as two-particle Breit \cite{CIMBPTSavukov1}, are not included in some codes, such as one used here. Cowan's code starts with non-relativistic Hartree-Fock approximation, and relativistic effects are included through the spin-orbit term. To improve accuracy of this approximation, scaling factors are used. In heavy atoms, this approach has drawbacks, such as semi-empirical procedure instead of {\it ab initio} calculations.

While heavy atoms have various issues of accurate treatment of valence-valence, valence-core interactions and relativistic effects, the accuracy requirement is exacerbated by strong mixing, resulting from high density of excited states. Mixing and order reversal can lead to difficulties in identification. Landé g-factors are quite valuable, since they are closely related to terms and can be used to evaluate the strength of interactions between states, mostly neighboring states with strong mixing. Transition probabilities which are often used to generate theoretical emission spectra are not known accurately. Moreover, observed and identified transitions in U and similar actinide atoms involve excited states that are more difficult for theory. Thus tests of atomic theory based on transition probabilities and lifetimes in the energy range were the theory should be more reliable  are quite limited.

Currently, the most important issue is addressed that 6 valence electrons would lead to a very large CI space, if the starting potential included only the core electrons. The starting potential was chosen to include $5f^4$ valence electrons, with the resulting potential +2, which is almost optimal, +1. However, CI-MBPT approach was not developed for including MBPT corrections properly in this potential, so they are set to zero, and the theory should be referred to as RCI. Here, the RCI method was applied to calculations of hyperfine constants. We performed calculations for low-energy states where single-configuration approximation can be quite accurate, so the mixing between configurations of different symmetry were not included, only the excitations of the same symmetry to correct single-electron orbitals. This considerably simplified the task, since the size of the Hamiltonian matrix was smaller than 20,000 by 20,000.

\section{RCI and CI-MBPT approach}
RCI method used here is implemented in the CI-MBPT code,   developed for open shell atoms with multiple valence electrons (see for example \cite{DzubaGe}), with MBPT corrections, to be defined later, set to zero. The theory can be summarized as follows. The effective CI+MBPT Hamiltonian for U I is split into two parts:
\begin{equation}
H^{eff}=\sum_{i=1}^{M} h_{1i}+\sum_{i\ne j}^{M} h_{2ij}.
\end{equation}
The one-electron contribution
\begin{equation}
 h_{1}=c\mathbf{\alpha}\cdot\mathbf{p}+(\beta-1)mc^2-Ze^2/r+V^{N-M}+\Sigma_1
 \end{equation}
in addition to the $V^{N-M}$ Dirac-Hartree-Fock (DHF) potential (here N is the total number of electrons and M is the number of removed valence electrons from the starting potential) contains the valence electron self-energy correction, $\Sigma_1$ \cite{Dzuba1987}.  In the current CI+MBPT program, the self-energy correction is calculated with the second-order MBPT. 
The two-electron Hamiltonian is
\begin{equation}
h_2=e^2/|\mathbf{r_1}-\mathbf{r_2}|+\Sigma_2
\end{equation}
where $\Sigma_2$ is the term accounting for Coulomb interaction screening arising from the presence of the core \cite{Dzuba89}. In the CI-MBPT program used, the screening is also calculated in the second order. Further details on the CI+MBPT approach can be found in Ref.\cite{Dzuba96}. The MBPT corrections can be correctly included for the case when all valence electrons are removed from the starting potential, that is when M=6 in case of U I.  

As we mentioned earlier, the starting $V^{N-6}$ potential is quite poor, and a large basis set is needed to correct it in valence-valence CI, but this would result in a very large Hamiltonian matrices. 

The ideal starting potential would be $V^{N-1}$, but it would be necessary to change it for different configurations. For example, $5f^36d7s^2$ configuration would need for each valence electron different starting potential: $5f^36d7s$ for the $7s$ electron, $5f^37s^2$ for $6d$, and so on. When calculating transitions that involve two different states, it becomes even more complicated. Because most configurations contain $5f^3$ and $5f^4$ parts, we chose the $5f^4$ starting potential, where one $5f$ electron roughly approximates the screening by other possible electrons $6d$ or $7s$.

We found that this starting potential leads to quite satisfactory results for most states we considered in this work, and some expansion over basis sets to be defined later was added to correct the difference between orbitals of the physical states and the lowest valence states in the basis sets.  

In terms of specific numerical steps, first, the DHF V$^{N-2}$ potential containing $5f^4$ valence electrons is calculated. Second, the basis in the frozen V$^{N-2}$ potential is calculated with the help of a B-spline subroutine for the ion in a cavity of radius $R=30$ a.u. The basis is then used to evaluate the CI+MBPT terms in Eq. 1,  with the MBPT part set to zero. Finally, the eigenvalue problem is solved for the effective Hamiltonian matrix (Eq.1).

Single-configuration RCI was tested and results were satisfactory. When the starting potential is not ideal ($5f^4$ instead of $f^37s^26d$ for the ground state), single excitations can be added to correct single-electron orbitals. For example, we added such excitations up to 15s for the 7s, 15p for 7p, 14d for 6d, and 13f for 5f states. Resulting matrices in RCI in some cases were as large as 20,000 by 20,000, but acceptable for available computing resources. Including double excitations or single excitations that would change  symmetry of configurations, for example,  the 5f to 7p single excitation, would lead to physical configuration mixing, and such excitations were excluded. As it has been shown \cite{Guyon72} many considered states classified as $f^3ds^2$ here are almost 100\% $f^3ds^2$ configuration pure. 
\begin{table}
\caption{ Land\'e g-factors for some low levels of different symmetry.  Subscripts 1 and vse mean RCI calculations based on single dominant configurations and a set of configurations generated by valence single excitations of this configuration, preserving orbitals' symmetry.    
  \label{TableLaIIJ1ev}}

  \begin{tabular}{rrlrrrrc}
  \hline
  \hline
E, cm$^{-1}$	&	J	& Conf.		&	g$_ {exp}$	&	g$_1$	& 	g$_{vse}$	&	$\sigma_{exp,vse}$	\\
\hline
0	&	6	&	$f^3ds^2$	&	0.750	&	0.7454	&	0.7391	&	1.5\%	\\
4275	&	6	&	$f^3ds^2$	&	0.920	&	0.9157	&	0.9138	&	0.7\%	\\
7005	&	6	&	$f^3ds^2$	&	0.950	&	0.9502	&	0.9534	&	0.4\%	\\
3800	&	7	&	$f^3ds^2$	&	0.925	&	0.9231	&	0.9199	&	0.6\%	\\
6249	&	6	&	$f^3d^2s$	&	0.625	&	0.592	&	0.589	&	6.1\%	\\
7326	&	7	&	$f^3ds^2$	&	1.020	&	1.0445	&	1.0479	&	2.7\%	\\
10069	&	7	&	$f^3ds^2$	&	0.930	&	0.9489	&	0.9357	&	0.6\%	\\

15631	&	7	&	$f^2d^2s^2$	&	0.910	&	0.9252	&	0.9227	&	1.4\%	\\
16900	&	7	&	$f^3dsp$	&	0.875	&	0.8364	&	0.837	&	4.5\%	\\
\hline
\hline
\end{tabular}

\end{table} 
\subsection{Land\'e g-factors: test of single-configuration approximation}
The first test was Land\'e g-factor calculations, using the RCI method described in the previous section, which are presented in Table I. As it can be seen, the agreement for considered states is quite accurate, which can serve as the first indication that the single-configuration approximation in the RCI framework is reasonable. Some disagreement can be observed for the $f^3d^2s$ state, with deviation as large as 6.1\%, but this can be attributed to mixing with some $f^3ds^2$ states. Inclusion of valence single excitations preserving configuration symmetry (vse) did not change much result, as expected, since such excitations mostly affect the radial orbitals and $g$-factors do not depend much on the radial part, only slightly due to relativistic effects, through deviation from the LS coupling scheme. 
\subsection{Lifetime calculation of the first $f^3dsp$ J=7 even state}
Lifetimes can be measured accurately, and they are used to derive transition probabilities from branching ratios. Thus lifetime errors can propagate to errors in transition probabilities. One additional issue for getting transition probabilities from lifetimes is that not all possible transitions are accounted in the experiment, so this can be a source of additional error.  Theory is better in this respect since it can generate a complete set of transitions, especially those outside the observable range. 
 Experimental U I lifetimes and transition probability data are given in most cases for relatively highly excited states, for which configuration mixing can be a problem; however, there is a state that is not strongly mixed: J=7 even $f^3dsp$ that has energy 16,900 cm$^{-1}$. We chose this state for our RCI calculations to test the theory. 
 
 The E1 transition probabilities $A$ are calculated from line strengths $S$:
\begin{equation}
A=\frac{2.142\times 10^{10}\omega^3 S} {2J+1}
\end{equation}
where $\omega$ is the experimental transition energy in atomic units, $J$ is the total angular momentum of the upper state. The lifetime was measured with three different methods: direct electron excitation \cite{dee} $\tau=205\pm 20$ ns, laser photo-ionization \cite{lp} $\tau=255\pm 25$ ns, and heavy ion sputtering \cite{his} $\tau=232\pm 25$ ns.
 The  average value is $231\pm25$ ns, while single-configuration theory gave 242 ns, in close agreement with experiment. Table II shows a breakdown of different decay channels, with a few very weak omitted. 
 Theoretical calculations that included valence single excitations (vse) gave somewhat longer lifetime, but it is expected that theory has significant uncertainty of this order, for example due to mixing of configurations of different symmetry. 
\begin{table}
\caption{Calculation of lifetime of the 16,900 cm$^{-1}$ $f^3dsp$ J=7 even state in single-configuration ``1'' and valence single-excitation ``vse'' approximations. The theoretical lifetimes $\tau_1$=1/A$^{tot}_1$=242 ns and $\tau_{vse}$=1/A$^{tot}_{vse}$=287 ns are in agreement with experimental lifetime $231\pm25$ ns within theoretical uncertainty that can be estimated from the difference $\tau_1-\tau_{vse}$. Energy is in cm$^{-1}$, vacuum wavelength in  \AA, and transition probabilities in s$^{-1}$. Square brackets denote powers of 10.      
  \label{TableLaIIJ1ev}}
  \begin{tabular}{rccccccc}
  \hline
  \hline
E$_{low}$	&$\lambda$&	Low Conf. 		& J$_{low}$		&	ME$_1$	&	A$_1$	&	ME$_{vse}$	&	A$_{vse}$	\\
\hline
0	&	5917.2	&	$f^3ds^2$	&	6	&	1.14	&	8.54[5]	&	1.07	&	7.40[5]	\\
3801	&	7634.2	&	$f^3ds^2$	&	7	&	0.96	&	2.81[5]	&	1.06	&	3.42[5]	\\
4275	&	7920.8	&	$f^3ds^2$	&	6	&	0.32	&	2.84[4]	&	0.31	&	2.69[4]	\\
6249	&	9388.8	&	$f^3d^2s$	&	6	&	1.16	&	2.19[5]	&	1.04	&	1.77[5]	\\
7326	&	10445.0	&	$f^3ds^2$		&	7	&	0.02	&	5.57[1]	&	0.04	&	1.47[2]	\\
7645	&	10805.0	&	$f^3ds^2$		&	8	&	0.13	&	1.71[3]	&	0.17	&	3.05[3]	\\
8118	&	11386.9	&	$f^3d^2s$		&	7	&	5.47	&	2.74[6]	&	4.89	&	2.19[6]	\\
10347	&	15260.2	&	$f^3d^2s$		&	8	&	0.27	&	2.77[3]	&	0.07	&	1.97[2]	\\
10685	&	16090.1	&	$f^3ds^2$	&	8	&	0.01	&	1.77[0]	&	0.01	&	3.04[0]	\\
\hline
Total	&		&		&			&		&	4.13[6]	&		&	3.48[6]	\\

\hline
\hline
\end{tabular}

\end{table}

\begin{table}
\caption{Magnetic dipole $A$ hyperfine constants (in MHz) for some low levels of different symmetry.  Experimental values are taken from \cite{AvrilTh} for the odd $f^3ds^2$ and $f^3d^2s$ states,  \cite{Avril} for the even $f^2d^2s^2$ state, and \cite{Hackel} for the even $f^3dsp$ state.   The deviation of theory from experiment is calculated as $\sigma_{exp,th}=|A_{exp}-A_{avr}|/A_{exp}$. The theoretical error is estimated as $\sigma_{th}=|A_1-A_{vse}|/2 A_{avr}$, where $A_{avr}=(A_1+A_{vse})/2$.
  \label{TableLaIIJ1ev}}

  \begin{tabular}{rrlrrrrrr}
  \hline
  \hline
  E	&	J	&	Conf.	&	A$_{exp}$	&	A$_1$	&	A$_{vse}$	& A$_{avr}$	&	$\sigma_{exp,th}$	&$\sigma_{th}$ 	\\
\hline
0	    &	6	&	$f^3ds^2$	&	-60.54	&	-69.17	&	-55.54	&	-62.35	&	3\%	&	11\% 	\\
4275	&	6	&	$f^3ds^2$	&	-59.13	&	-47.58	&	-50.31	&	-48.94	&	17\%	&	3\%	\\
7005	&	6	&	$f^3ds^2$	&	-54.39	&	-45.14	&	-49.99	&	-47.56	&	13\%	&	5\%	\\
3800	&	7	&	$f^3ds^2$	&	-56.31	&	-44.30	&	-47.78	&	-46.04	&	18\%	&	4\%	\\
7326	&	7	&	$f^3ds^2$	&	-63.15	&	-30.59	&	-44.91	&	-37.75	&	40\%	&	19\%	\\
10069	&	7	&   $f^3ds^2$	&	-53.01	&	-44.24	&	-48.45	&	-46.35	&	13\%	&	5\%	\\
6249	&	6	&	$f^3d^2s$	&	67.8	&	89.43	&	90.14	&	89.79	&	32\%	&	0.4\%	\\

15631	&	7	&	$f^2d^2s^2$	&	-84.72	&	-42.98	&	-55.88	&	-49.43	&	42\%	&	13\%	\\
16900	&	7	&	$f^3dsp$	&	-171	&	-67.73	&	-82.58	&	-75.15	&	56\%	&	10\%	\\
\hline
\hline
\end{tabular}
\label{Ahfs}
\end{table}

\begin{table}
\caption{Magnetic dipole $A$ hyperfine constants (in MHz) for some low levels of different symmetry.  Experimental values are taken from \cite{AvrilTh} for the odd $f^3ds^2$ and $f^3d^2s$ states,  \cite{Avril} for the even $f^2d^2s^2$ state, and \cite{Hackel} for the even $f^3dsp$ state. The subscript ``vse,adj'' denotes RCI calculations based on a set of configurations generated by single-electron excitation of the dominant non-relativistic configuration preserving orbital symmetry with the result multiplied by 1.124.  The deviation of theory from experiment is calculated as $\sigma_{exp,th}=|A_{exp}-A_{vse,adj}|/A_{exp}$. The theoretical error is estimated by $\sigma_{th}=|A_1-A_{vse}|/2 A_{avr}$, the same as in the previous table, where $A_{avr}=(A_1+A_{vse})/2$.
  \label{TableLaIIJ1ev}}

  \begin{tabular}{rrlrrrr}
  \hline
  \hline
E	&	J	&	Conf.	&	A$_{exp}$	&	A$_{vse,adj}$	 & $\sigma_{exp,th}$ &$\sigma_{th}$	\\
\hline
0	    &	6	&	$f^3ds^2$	&	-60.54	&	-62.43	&	3\% & 11\%	\\
4275	&	6	&	$f^3ds^2$	&	-59.13	&	-56.55	&	4\%	& 3\%\\
7005	&	6	&	$f^3ds^2$	&	-54.39	&	-56.19	&	3\%&	5\%	\\
3800	&	7	&	$f^3ds^2$	&	-56.31	&	-53.70	&	5\%	&	4\%\\
7326	&	7	&	$f^3ds^2$	&	-63.15	&	-50.48	&	20\%&	19\%	\\
10069	&	7	&	 $f^3ds^2$	&	-53.01	&	-54.46	&	3\%&	5\%	\\
6249	&	6	&	$f^3d^2s$	&	67.8	&	101.32	&	49\%&0.4\%	\\
15631	&	7	&	$f^2d^2s^2$	&	-84.72	&	-62.81	&	26\%&	13\%	\\
16900	&	7	&	$f^3dsp$	&	-171	&	-92.82	&	46\%&	10\%	\\
\hline
\hline
\end{tabular}
\label{Ahfs2}
\end{table}

\subsection{RCI calculations of magnetic-dipole $A$ hyperfine-structure constants}
Calculations of magnetic dipole hyperfine structure $A$ constants are performed with  a single-configuration and vse 
RCI (Table \ref{Ahfs}). The two methods were used to estimate the theoretical uncertainty, $\sigma_{th}=|A_{vse}-A_1|/(A_{vse}+A_1)$. The magnetic gyromagnetic ratio, which is the input of the RCI program, of U-235 was initially taken $-0.10857$,  which was calculated from $\mu=-0.38(3)\mu_N$ \cite{PRLmuU235}, using well-known relation:  $g_I=\mu/\mu_N/I=-0.38/(7/2)$. This value was derived from hyperfine structure measurements using Dirac Hartree-Fock calculations, so it is not direct measurement and is the subject to uncertainty of the theoretical interpretation. This result is in agreement with older determinations with electron-nuclear double resonance (-0.36) and electron paramagnetic resonance (-0.35)\cite{oldisotopetable}. 

It can be noted that many RCI single-configuration hyperfine constants $A$ of the considered odd states are in good agreement with experiment. These states can be accurately characterized by a single non-relativistic configuration, $f^3ds^2$, as it is given in Table I of \cite{AvrilTh}. We also obtained a good agreement for the $g$-factors of these states, which is expected for pure configurations.  When the state is characterized by only one non-relativistic configuration, it is still important to know the distribution over different terms or relativistic configurations formed from this non-relativistic configuration. Then correct inclusion of relativistic effects, as it is done in RCI, is essential for accurate {\it ab initio} calculations. When we compare un-scaled average $A_{avr}=(A_1+A_{vse})/2$ values with the experiment (Table \ref{Ahfs}), the deviations from experiment are poorly correlated with theoretical uncertainty, so it is quite possible that a systematic shift is present, due to error in the value of the magnetic moment of U-235.

The vse model gives systematically better agreement with experiment than the single-configuration model. Moreover, if the vse model result is multiplied by a close to unity factor 1.124, the agreement with experiment of the vse model for the four lowest and the sixth levels becomes within 5\%, Table~\ref{Ahfs2}, and agreement for the all shown states except for the $f^2d^2s$ state is improved compared to the $A_{avr}$ or $A_1$ in Table \ref{Ahfs}. In cases where mixing is not strong, the theoretical uncertainty, $\sigma_{th}$, is correlated with the deviations from experiment, $\sigma_{exp,th}$, but in cases where mixing is strong, such as the case of $f^3d^2s$ state, it is not. This can be attributed to limited accuracy of the single-configuration approximation. 

While this result might need further investigation, there is possibility that the value 0.38(3) might be somewhat too low, and instead 0.43(2) can be recommended. This is quite possible because the new value is almost within error bar of the measurement, and the two values overlap within the combined theoretical and experimental error bars. The new value is actually in better agreement with another experiment, 0.46(3) \cite{JPC}. The measurement was done for trivalent U-235 in several crystals and $1/r^3$ value needed for the extraction of $\mu$ was taken from calculations. Here again the theoretical interpretation plays an important role for obtaining $\mu$ from the hyperfine splitting measurements, which themselves are quite accurate. 

\begin{table}
\caption{ Electric-quadrupole B hyperfine constants for some low levels of different symmetry.  Experimental values are taken from \cite{AvrilTh} for the odd $f^3ds^2$ and $f^3d^2s$ states,  \cite{Avril} for the even $f^2d^2s^2$ state, and \cite{Hackel} for the even $f^3dsp$ state. Subscripts 1 and vse mean RCI calculations based on a single dominant configurations and a set of configurations generated by single-electron excitation of this configuration preserving orbital symmetry. The deviation of theory from experiment is calculated as $\sigma_{exp,th}=|B_{exp}-B_{avr}|/B_{exp}$. Theoretical uncertainty is estimated as $(B_1-B_{avr})/B_{avr}$, where $B_{avr}=(B_1+B_{vse})/2$ 
  \label{TableLaIIJ1ev}}

  \begin{tabular}{rrlrrrrrr}
  \hline
  \hline
E, cm$^{-1}$	&	J	&	Conf.	&	B$_{exp}$	&	B$_1$	&	B$_{vse}$	& B$_{avr}$	&	$\sigma_{exp,th}$	&$\sigma_{th}$	\\
\hline
0	&	6	&	$f^3ds^2$	&	4074	&	2884	&	4242	&	3563	&	13\%	&	19\%	\\
4275	&	6	&	$f^3ds^2$	&	447	&	795	&	649	&	722	&	62\%	&	10\%	\\
7005	&	6	&	$f^3ds^2$	&	609	&	126	&	483	&	304	&	50\%	&	59\%	\\
3800	&	7	&	$f^3ds^2$	&	4112	&	3082	&	4362	&	3722	&	9\%	&	17\%	\\
7326	&	7	&	$f^3ds^2$	&	1239	&	1201	&	1066	&	1134	&	9\%	&	6\%	\\
10069	&	7	&	$f^3ds^2$	&	2744	&	2379	&	3002	&	2691	&	2\%	&	12\%	\\
6249	&	6	&	$f^3d^2s$	&	2366	&	1681	&	2099	&	1890	&	20\%	&	11\%	\\

15631	&	7	&	$f^2d^2s^2$	&	4422	&	4197	&	5568	&	4882	&	10\%	&	14\%	\\
16900	&	7	&	$f^3dsp$	&	2685	&	4087	&	5029	&	4558	&	70\%	&	10\%	\\
\hline
\hline
\end{tabular}
\label{Bhfs}
\end{table}

\begin{table}
\caption{ Electric-quadrupole B hyperfine constants for some low levels of different symmetry.  Experimental values are taken from \cite{AvrilTh} for the odd $f^3ds^2$ and $f^3d^2s$ states,  \cite{Avril} for the even $f^2d^2s^2$ state, and \cite{Hackel} for the even $f^3dsp$ state. The theoretical B values ($B_{fit}$) obtained using fit for two 22 levels with 6 parameters are taken from \cite{AvrilTh}. Subscripts 1 and vse mean RCI calculations based on a single dominant configurations and a set of configurations generated by single-electron excitation of this configuration preserving orbital symmetry. The deviation of theory from experiment is calculated as $\sigma_{exp,th}=|B_{exp}-B_{vse}|/B_{exp}$. Theoretical uncertainty is estimated as $\sigma_{th}=(B_1-B_{avr})/B_{avr}$, where $B_{avr}=(B_1+B_{vse})/2$; $\sigma_{exp,th}=|B_{exp}-B_{vse}|/B_{exp}$ denotes the deviation of vse theory from experiment. 
  \label{TableLaIIJ1ev}}

  \begin{tabular}{rrlrrrrrr}
  \hline
  \hline
E, cm$^{-1}$	&	J	&	Conf.	&	B$_{exp}$ & $B_{fit}$		&	B$_{vse}$	&	$\sigma_{exp,th}$	&$\sigma_{th}$	\\
\hline
0	&	6	&	$f^3ds^2$	&	4074	&4049&	4242	&	4\%	&	19\%	\\
4275	&	6	&	$f^3ds^2$	&	447	&633&	649  	&	45\%	&	10\%	\\
7005	&	6	&	$f^3ds^2$	&	609	&303&	483 	&	21\%	&	59\%	\\
3800	&	7	&	$f^3ds^2$	&	4112&4148&	4362	&	6\%	&	17\%	\\
7326	&	7	&	$f^3ds^2$	&	1239&1424&	1066	&	14\%	&	6\%	\\
10069	&	7	&	$f^3ds^2$	&	2744&2934&	3002	&	9\%	&	12\%	\\
6249	&	6	&	$f^3d^2s$	&	2366&&	2099	&	11\%&	11\%	\\

15631	&	7	&	$f^2d^2s^2$	&	4422&&	5568	&	26\%&	14\%	\\
16900	&	7	&	$f^3dsp$	&	2685&&	5029	&	87\%&	10\%	\\
\hline
\hline
\end{tabular}
\label{Bhfs2}
\end{table}

\subsection{RCI calculations of electric-quadrupole B hyperfine structure constants}
In case of the quadrupole constant calculations we took $Q=+4.55(9)$ from Mu-X measurement method \cite{oldisotopetable}. 
RCI calculations of the electric-quadrupole B hyperfine constant were performed with the two methods, similarly to the $A$ constant. Table\ref{Bhfs} shows the comparison of the two methods with experiment, with the relative difference between $B_{avr}$ and $B_{exp}$ explicitly given as the measure of deviation of theory from the experiment. Table\ref{Bhfs2} shows the relative deviation of vse calculations from experiment, which appear more accurate for low-energy states than $B_{avr}$, where the assumption of single configuration approximation can be justified.  In case of the both models, agreement between theory and experiment for lowest states considered here, which have $B$ values large, was close; however, for small $B$ values, the agreement was poor. The vse $B$ values in general agree better with experiment than the single-configuration values. The highest considered level (16,900 cm$^{-1}$) has poor agreement, which as in case of $A$ values, can be attributed to departure from the single-configuration approximation, according to the $g$-factor. In contrast to $A$ calculations, there is no obvious need for rescaling $Q$. 

\section{Discussion}
In \cite{AvrilTh} it has been stated that odd 23 levels for which experimental A and B values were available belong to the $5f^36d7s^2$ configuration with purity higher than 97.5\%. This was not confirmed with the observed isotopic shifts, in particular for the odd 10,069 cm$^{-1}$ level considered in the current work \cite{BlaiseTable}. Nevertheless, our calculation of A constant for this level indicates close agreement with experiment, implying validity of the single-configuration assumption. The hyperfine structure calculations of \cite{AvrilTh} were based on fine structure parameters given by Guyon \cite{Guyon}. In order to take into account relativistic effects with non-relativistic approach the authors of \cite{AvrilTh} used the formalism of tensor operators of Judd et al.\cite{Judd} and Sandars and Beck \cite{SandarsBeck}. They obtained quite accurate agreement (sometimes a few \%) with the fit of 6 parameters for A values and 6 parameters for B values for many low-energy $5f^36d7s$ states with J from 1 to 9. In addition, a number of hyperfine constants for un-observed states were predicted. However, their {\it ab initio} results were much less accurate. Relativistic effects taken into account with 6 coefficients for A values were quite significant. 
Our relativistic calculations without adjustable parameters, except maybe one introduced to correct possible error in the assumed magnetic moment of U-235, gives a few percent accuracy for lowest states, including that of 10,069 cm$^{-1}$. We also performed calculations beyond $5f^36d7s^2$ states. The accuracy was not as high, and it can be attributed to in-applicability of the single-configuration approximation. The work \cite{AvrilTh} does not provide calculations for other than $5f^36d7s$ states. 
Work of \cite{AvrilTh} also presented results for B hyperfine structure constants, assuming $Q=+4.55$. For low-energy states which have small B values, the accuracy was poor as in our calculations, especially when the fit involved 22 levels. The accuracy was somewhat improved when the number of fitted levels was co-measurable with the number of fitted parameters: 8 to 6. It is also interesting to note that values obtained through the fit of 22 levels is close to our theoretical values Table\ref{Bhfs2}: for 4275 level, our $B_1=795$ and $B_1=649$ MHz while \cite{AvrilTh} gives $B=633$ MHz; for 7005 state, our $B_1=126$ and $B_1=483$ MHz, while \cite{AvrilTh} gives $B=303$ MHz.
It can be concluded that our vse model calculations and semi-empirical method of \cite{AvrilTh} give quite accurate results when single-configuration approximation is valid and the B values are not small, but the calculations have to be refined for other cases, which we plan to do in the future.  

\section{Conclusions}
In conclusion, the studies of U-235 hyperfine constants are important for testing theory of complex atoms and have applications in nuclear forensics. We found that RCI gave g-factors in close agreement with experiment, justifying the single-configuration approximation. RCI method reproduced experimental hyperfine A and B constants quite accurately for low-energy states where mixing is small, and in case of $B$ when the $B$ value was not small. We also found that a scaling of $\mu$ with a factor of 1.124, considerably improves vse calculations for $A$ hyperfine constants, so slightly larger $\mu=0.43(2)$ can be recommended. RCI also gave reasonable value of lifetime of relatively high level. Thus it can be concluded that the RCI method discussed here is quite promising for calculations of complex atoms, such as U I, and for applications in fundamental physics.

\section{Acknowledgement} The work has been performed under the auspices of the U.S. DOE by LANL under contract No. DE-AC52-06NA25396. The author is grateful to Dr. Dzuba for making his CI-MBPT code available for this work.

\end{document}